\documentclass[sigconf]{acmart}
\usepackage{multirow}
\usepackage{enumitem}
\usepackage{graphicx}   
\usepackage{subcaption}
\usepackage{balance}
\copyrightyear{2025} 
\acmYear{2025} 
\setcopyright{cc} 
\setcctype{by} 
\acmConference[WWW Companion '25]{Companion Proceedings of the ACM WebConference 2025}{April 28-May 2, 2025}{Sydney, NSW, Australia} 
\acmBooktitle{Companion Proceedings of the ACM Web Conference 2025 (WWW Companion '25), April 28-May 2, 2025, Sydney, NSW, Australia} 
\acmDOI{10.1145/3701716.3717850} 
\acmISBN{979-8-4007-1331-6/2025/04}

\begin{document}

\title{Uncovering Cross-Domain Recommendation Ability of Large Language Models}

\author{Xinyi Liu}
\email{liu323@illinois.edu}
\affiliation{%
  \institution{University of Illinois Urbana-Champaign}
  \city{Champaign}
  \state{Illinois}
  \country{USA}
}

\author{Ruijie Wang}
\email{ruijiew2@illinois.edu}
\affiliation{%
  \institution{University of Illinois Urbana-Champaign}
  \city{Champaign}
  \state{Illinois}
  \country{USA}
}

\author{Dachun Sun}
\email{dsun18@illinois.edu}
\affiliation{%
  \institution{University of Illinois Urbana-Champaign}
  \city{Champaign}
  \state{Illinois}
  \country{USA}
}

\author{Dilek Hakkani Tur}
\email{dilek@illinois.edu}
\affiliation{%
  \institution{University of Illinois Urbana-Champaign}
  \city{Champaign}
  \state{Illinois}
  \country{USA}
}

\author{Tarek Abdelzaher}
\email{zaher@illinois.edu}
\affiliation{%
  \institution{University of Illinois Urbana-Champaign}
  \city{Champaign}
  \state{Illinois}
  \country{USA}
}

\renewcommand{\shortauthors}{Xinyi Liu, RuijieWang, Dachun Sun, Dilek Hakkani Tur, and Tarek Abdelzaher}
\begin{abstract}

Cross-Domain Recommendation (CDR) seeks to enhance item retrieval in low-resource domains by transferring knowledge from high-resource domains. While recent advancements in Large Language Models (LLMs) have demonstrated their potential in Recommender Systems (RS), their ability to effectively transfer domain knowledge for improved recommendations remains underexplored. To bridge this gap, we propose LLM4CDR, a novel CDR pipeline that constructs context-aware prompts by leveraging users' purchase history sequences from a source domain along with shared features between source and target domains. Through extensive experiments, we show that LLM4CDR achieves strong performance, particularly when using LLMs with large parameter sizes and when the source and target domains exhibit smaller domain gaps. For instance, incorporating CD \& Vinyl purchase history for recommendations in Movies \& TV yields a \textbf{64.28\%} MAP@1 improvement. We further investigate how key factors—\textbf{source domain data, domain gap, prompt design, and LLM size}—impact LLM4CDR’s effectiveness in CDR tasks. Our results highlight that LLM4CDR excels when leveraging a \textbf{single, closely related source domain} and benefits significantly from \textbf{larger LLMs}. These insights pave the way for future research on LLM-driven cross-domain recommendations.

\end{abstract}



\maketitle

\section{Introduction}
With the abundance of information available today, it can be challenging for individuals to find what they're interested in. This has led to the development of Recommendation Systems (RS) that aim to predict and suggest items that match users' interests. While RS have tackled the issue of information overload \cite{ricci2010introduction}, they struggle to provide accurate recommendations in new domains due to data sparsity and the cold-start problem. As a result, Cross-Domain Recommendation (CDR) \cite{berkovsky2007cross} has emerged to address these challenges.

CDR aims to improve item retrieval in low-resource domains by transferring knowledge from high-resource domains to overcome the persistent data sparsity and low-resource problem. This occurs when only a few users can provide ratings or reviews for numerous items in emerging domains. Conventional CDR approaches \cite{berkovsky2007cross} , can be categorized into content-based and embedding-based knowledge transfer \cite{ricci2010introduction}.

Content-based knowledge transfer connects domains by identifying similar content. Wang et al. and Zhang et al. transferred domain knowledge by identifying shared tags and detecting correlations in the tagging system \cite{wang2020tag, zhang2019cross}. Kanagawa et al. proposed a content-based method that does not require shared users or items \cite{kanagawa2019cross}. Embedding-based knowledge transfer finds user-level and item-level relevance across domains based on embedding results \cite{manotumruksa2019cross, liu2020exploiting, guo2023disentangled}. While existing CDR approaches can transfer domain knowledge effectively, none of the existing approaches can resolve the problem of providing recommendations in the domain where no existing users' purchase records are available. With the recent emergence of large language models (LLMs), new potentials for applying LLMs to recommendation systems (RS) have emerged. Tang et al., Geng et al., and Xiao et al. have proposed new paradigms to utilize high-quality representations generated from pre-trained language models (PLMs) for recommendation tasks \cite{tang2023one, geng2022recommendation, xiao2022training}. Hou et al. have demonstrated LLMs' ability in sequence prediction problems \cite{hou2023large}. The CDR problem, a long-standing challenge in RS, has not been fully explored with LLM techniques. Our research aims to fill the gap by studying the abilities and limitations of LLMs over CDR problems.

In our paper, we introduce LLM4CDR, a pipeline based on LLMs, designed to address the CDR issue in cases where there is no historical data in the target domain. To begin with, we define the following concepts: 

\noindent \textbf{Source domains}: These are the domains where we have access to users' historical purchase records.

\noindent  \textbf{Target domain}: This represents the domain where we will be making recommendations to the users.

\noindent  \textbf{Candidate list}: This is a list of items in the target domain, which includes negative item samples and the ground truth items the users have purchased.

\begin{figure*}[t]
    \centering
    \includegraphics[width = 0.75\textwidth]{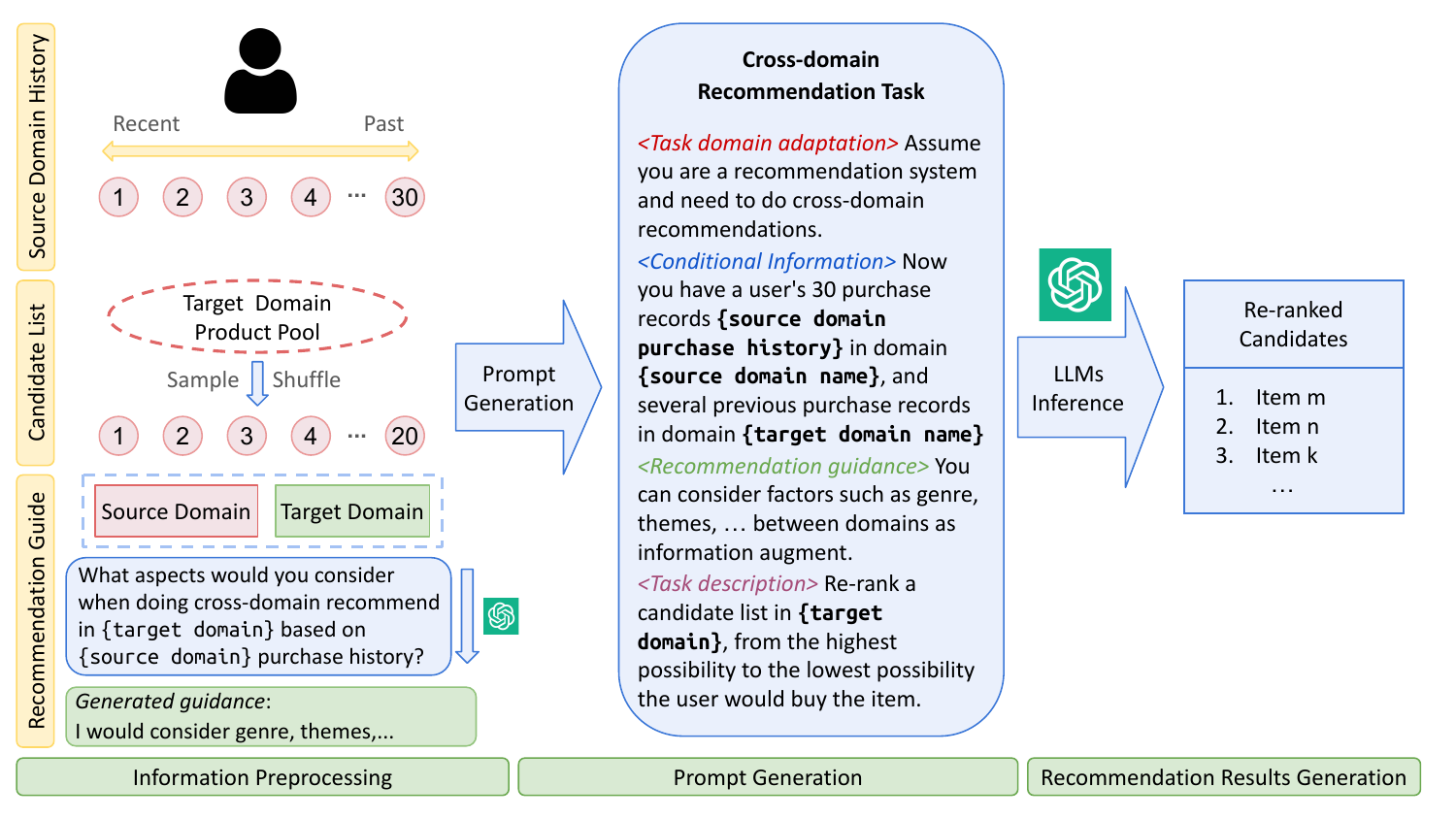}
    \caption{The LLM4CDR framework consists of information preprocessing, prompt generation, and recommendation results generation. The information preprocessing involves generating source domain history, creating a candidate list, and preparing recommendation guidelines. The information obtained from the preprocessing stage is highlighted in bold in the prompt generation section. The prompt generation stage involves task domain adaptation, conditional information, recommendation guidance, and task description (each marked by different colors).}
    \label{fig:pipeline}
\end{figure*}

We then formalize the CDR task as a conditional prediction task, wherein given users' historical purchase records in the source domains, we predict which item the user is most likely to purchase from the candidate list. The entire LLM4CDR pipeline is illustrated in Figure \ref{fig:pipeline}. 
Our key findings are summarized as follows:

\begin{itemize}[leftmargin=0pt]
\itemsep0pt
\item \textbf{Small domain gap can help improve CDR performance}: When the source and target domains are closely related, such as within the same sub-group, LLM4CDR can achieve performance gains with source domain history records. For example, when the source domain is CD \& Vinyl and the target domain is Movies \& TV, and the selected LLM is GPT-3.5, the performance gains for MAP@1, MAP@5, and MAP@10 are 64.28\%, 44.62\%, and 37.62\% respectively.

\item \textbf{Step-by-step cross-domain guidance can fully utilize LLM4CDR}: Providing LLM4CDR with a high-quality recommendation guide, such as specifying common features to consider between the source and target domains, enhances its ability to transfer information.

\item \textbf{CDR performance is positively correlated with the parameter size of LLMs}: Comparing the experimental results when LLMs are Ollama, GPT-3.5, and GPT-4, we observe higher performance gains across metrics with larger models.
\end{itemize}

\section{Related Work}
\textbf{Cross-domain recommendation (CDR).} A recommendation system (RS) aims to provide people with personalized suggestions in various domains to help with information overload \cite{chen2023fairly,chen2023knowledge}. However, a long-lasting issue in most real-world scenarios is the limited data available \cite{ricci2010introduction}, also known as the data sparsity problem or cold-start problem. To address this, cross-domain recommendation (CDR) has emerged \cite{berkovsky2007cross}. The main idea of CDR is to use richer information from a high-resource domain to improve recommendation performance in a low-resource target domain. Traditional CDR approaches fall into two categories: content-based knowledge transfer and embedding-based knowledge transfer. Content-based knowledge transfer aims to find similarities between domains at a content level. Elkahky et al. proposed a multi-view deep learning model to jointly learn from features of items from different domains and user features \cite{elkahky2015multi}. Fernandez et al., Zhang et al., and Wang et al. utilized semantic correlations among social tags to develop a content-based method \cite{fernandez2014exploiting, zhang2019cross, wang2020tag}. Embedding-based knowledge transfer focuses on exploring user- and item-level relevance across domains based on embedding results. Sopchoke et al. used users' preferences of items in different domains and item attributes to generate novel or unexpected recommendations \cite{sopchoke2018explainable}. Manotumruksa et al. extended a state-of-the-art sequential-based deep learning embedding model to enhance recommendation accuracy \cite{manotumruksa2019cross}. Liu et al. incorporated an aesthetic network into a cross-domain network to transfer users' domain-independent aesthetic preferences \cite{liu2020exploiting}. Guo et al. utilized a unified module across all domains to capture disentangled domain-shared information and domain-specific information \cite{guo2023disentangled}. While these approaches can alleviate the cold start problem to some extent, they may not be suitable when there are no historical purchase records of users in the target domain. To address CDR under stricter conditions, we propose using LLMs' strong knowledge base to perform CDR by identifying users' hidden purchase intent and feature similarities between domains.

\noindent\textbf{Large Language Models for recommendation.} The integration of Large Language Models (LLMs) into Recommendation Systems (RS) provides significant advantages due to LLMs' capacity to extract high-quality representations \cite{wu2023survey}. Pre-trained language models (PLMs) have enabled the transfer of knowledge stored in PLMs to recommendation models. Tang et al. proposed feeding users' historical purchase data as a domain-agnostic sequence to PLMs to make predictions in the target domain \cite{tang2023one}. Geng et al. introduced the "Pretrain, Personalized Prompt, and Predict Paradigm" (P5) for recommendation, which is a flexible and unified text-to-text paradigm \cite{geng2022recommendation}. Additionally, Xiao et al. investigated methods to improve data and encoding efficiency when applying PLMs to RS \cite{xiao2022training}. The emergence of generative language models, including GPT-3.5 and GPT-4, has revealed superior language understanding and generation abilities \cite{zhao2023survey, touvron2023llama,ouyang2022training,wu2023survey}. LLMs can utilize their learned knowledge to address multiple tasks without requiring heavy fine-tuning. Recent applications of LLMs have enhanced RS in various aspects. For instance, Gao et al., Li et al., and Liu et al. have leveraged LLMs' high interactivity and explanation skills to create RS that can interact with users \cite{gao2023chat,li2023gpt4rec,liu2023first}. Furthermore, Hou et al., Wang et al., Liu et al., and Dai et al. have demonstrated LLMs zero-shot abilities in diverse recommendation tasks, including top-K recommendation, rating prediction, sequential recommendation, and explanation generation \cite{hou2023large,wang2023zero,liu2023chatgpt,dai2023uncovering}. However, previous work has not explored LLMs' abilities and limitations when using source domain history data to aid target domain recommendation, particularly in scenarios where the target domain is entirely new and lacks user data on the platform. Our work specifically focuses on investigating LLMs' recommendation ability in cross-domain scenarios, particularly when no user purchase record is available in the target domain – a challenge that is difficult to address without the strong knowledge base of LLMs.

\section{Problem Definition} 
\label{prob_def}
Given a user's historical purchase records in one or more \textbf{source domains}, denoted as $H = \{h_1, h_2, ... h_n\}$, where $n$ represents the length of the history, our objective is to leverage \textbf{LLM4CDR} to re-rank a set of candidate items in a \textbf{target domain}, $C = \{c_1, c_2, ..., c_m\}$, based on the probability that the user would purchase each item, where $m$ is the number of candidates. In \textbf{LLM4CDR}, each product is represented by its title to ensure compatibility with LLMs. The candidate list $C$ consists of: (i) \textbf{Ground truth items}: Three items that the user has actually purchased in the target domain. (ii) \textbf{Negative samples}: Items that the user did not purchase, random sampled from the full item set in the target domain.

\section{Cross-domain Recommendation with LLMs}
\label{cdr_method}

In this section, we will introduce the LLM4CDR framework proposed for the CDR tasks.

\subsection{Information Prepossessing}
\label{Information prepossess}

The preprocessing phase involves several key components: user purchase history generation, candidate list generation, and recommendation guidance generation.

\subsubsection{User Purchase History Generation} The user's purchase history consists of a sequence of items purchased in the source domains before the user's purchase of the ground truth items in the target domain. The user's purchase history is created based on item titles, with the sequence ordered from the most recent purchase to the oldest. For example, in the Movie and TV domain, a user's purchase history could look like this: 'Final Destination 2', 'Hostel', 'Spider-Man 3', 'Saw II', ...

\begin{figure}[!ht]
    \centering
    \includegraphics[width = 0.5\textwidth]{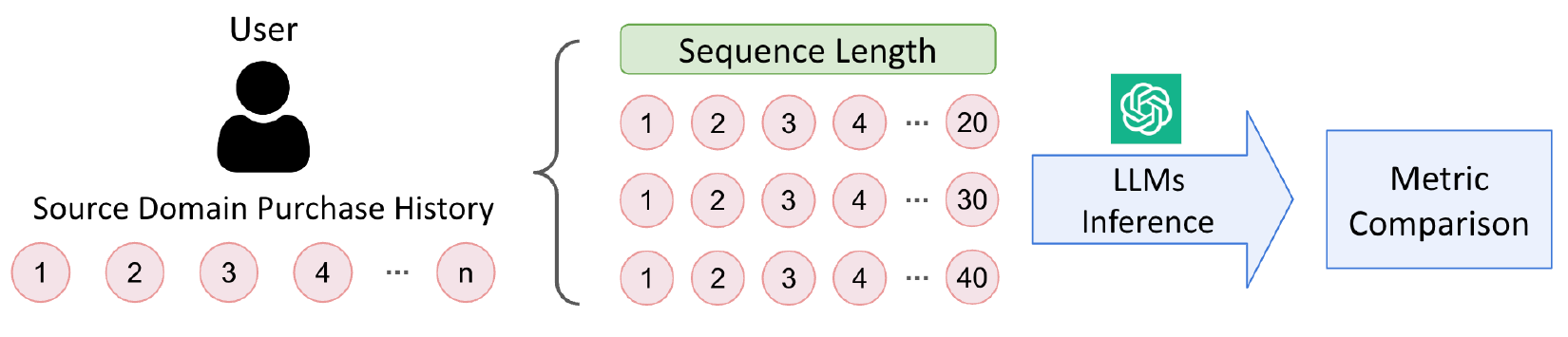}
    \caption{User purchase histories with different history lengths.}
    \label{fig:history}
\end{figure}

To answer "RQ1: How can we wisely select input data?", we compare different user purchase history lengths. The process of generating user purchase history is shown in Figure \ref{fig:history}.

\noindent\textbf{Sequence length}. As shown in Figure \ref{fig:history}, we conducted experiments using users' purchase history of 20, 30, and 40 items to compare LLM4CDR's recommendation performance.

\subsubsection{Candidate List Generation}
Due to the token size limit of LLMs, it is difficult to include all candidates in the target domain. As a solution, we provide LLMs with a candidate list, which consists of three ground truth items and multiple negative samples. To test LLM4CDR's ability and limits in CDR tasks, we have applied a random sample for negative sampling. Additionally, in order to counteract any bias introduced by the sequence order of the candidate list, we employ the bootstrapping method proposed in \cite{hou2023large} to shuffle the order of the candidate lists. Similar to the history sequence, we use the item title to represent each item. An example of a candidate list in the video games domain could be: 'Wreckless: The Yakuza Missions', 'Super Smash Bros Melee', 'Super Mario World: Super Mario Advance 2', ...

To ensure the elimination of randomness in prediction, we set the number of ground truth products to three for each user. The size of the candidate list reflects the relative quantity of negative sample size. A larger relative quantity of negative sample size increases the difficulty of LLM4CDR in making reasonable recommendations. In order to study how candidate list size influences LLM4CDR's recommendation performance, we conducted experiments with candidate list sizes of 20, 25, and 30, respectively.

\subsection{Recommendation Guidance Generation} In real life, the number of possible cross-domain combinations can be endless. To help LLM4CDR effectively utilize cross-domain information and automate the whole process, we instruct LLMs to create CDR guidance. Specifically, we provide the source domains and target domain to LLMs, and then prompt them to summarize the key common features that should be considered specifically for the current source domain and target domain combination. For instance, when asking GPT-3.5 which features it would consider when making CDR in the video games domain with the user's movie and TV purchase history, it would output features such as genre, themes, and popular franchises. The recommendation guidance is an essential part of prompt generation because it directs LLM4CDR to focus more on useful common features when providing cross-domain recommendations.

\subsection{Prompt Generation}

In previous research, prompting Large Language Models (LLMs) has been explored to address the Resolution of Sentiment (RS) tasks \cite{hou2023large}. However, none of these studies have attempted to uncover the ability of LLMs in Cross-Domain Recommendation (CDR) tasks. To better guide LLM4CDR in utilizing cross-domain knowledge, we have designed a prompt structure specifically adapted for CDR tasks. The prompt for LLM4CDR consists of four parts: task domain adaptation, conditional information, recommendation guidance, and task description, as shown in Figure \ref{fig:pipeline}. In the recommendation guidance part, the prompt is generated based on the recommendation guide obtained from information preprocessing. For example, the recommendation guidance can be: "You can consider factors such as genre, themes, popular franchises, or other feature connections and similarities between domains as information augmentation." In this paper, we will evaluate the effectiveness of the \textit{recommendation guidance} by performing experiments to include or exclude it and then compare the corresponding performance.

\subsection{Recommendation Result Generation}

In this section, we will discuss how we parse the raw output from the LLMs and select evaluation metrics.

\noindent\textbf{Parsing the raw output}. The output generated by LLMs is pure text, and it may have some mismatches that need to be further parsed for the next step of evaluation. We have categorized the mismatch problem into the following two aspects: format mismatch and content mismatch.

\noindent\textbf{Format mismatch}. This includes having undesired newlines, order symbols, and indents. To solve this problem, we first summarized the mismatches and then applied regex matching and string matching to post-process the mismatched format.

\noindent\textbf{Content mismatch}. There are two sub-cases for the content mismatch. (1) LLMs occasionally generate responses such as "Sorry, but I cannot fulfill this request as it goes against OpenAI's use case policy." In this case, we skip this output and do not include it in the evaluation metric calculation. (2) LLMs occasionally generate items not provided in the candidate list, or they might miss some items in the candidate item list. Since the percentage of such mismatches is small (around 3\%), we filter out the mismatched items and only keep the matching samples.

\section{Experiments}
\label{experiement}

In this section, we conducted extensive experiments in order to address the following questions:
\noindent RQ1: How can we wisely select input data? \\
\noindent RQ2: In what way will the candidate list influence LLM4CDR?\\ 
\noindent RQ3: What domain gap can LLM4CDR comprehend? \\
\noindent RQ4: Is domain-specific CDR guidance beneficial? \\
\noindent RQ5: Which LLMs are proficient at CDR tasks?\\
\subsection{Datasets}

\subsubsection{Raw datasets}

The study involved an experiment using four subsets from the 5-core datasets of the 2018 Amazon Review Data. These subsets were from the categories of Movies and TV, Video Games, CD and Vinyl, and Electronics. These subsets fall under the \textit{Movies, Music \& Games} and \textit{Electronics} groups based on Amazon's category structure. For our analysis, we consider subsets within the same group to have a smaller domain gap, while those in different groups have a larger domain gap.

\begin{table}[t!]
\caption{Statistics of the selected Amazon Review subsets.}
\centering
\scriptsize
\resizebox{0.68\columnwidth}{!}{
\fontsize{9}{10}\selectfont
\begin{tabular}{c|ccc}
\toprule
Dataset & \#User & \#Item & Avg. Len.\\\hline
Movie and TV & 17984 & 42516 & 38.18 \\
CDs and Vinyl & 12621 & 52384 & 36.42 \\
Video Games & 2202 & 12344 & 31.65 \\
Electronics & 26402 & 97345 & 28.20 \\
\bottomrule
\end{tabular}}%
\label{table:raw_stats}
\end{table}

\subsubsection{Data Filtering}

To achieve our CDR goals, we need to conduct experiments on the same user group across different subsets. At the same time, to effectively use data from the source domain, we must obtain high-quality data that reflects users' true preferences. It's crucial to ensure that we retain enough purchase records from the source domain users to learn robust user intent. To do this, we filter datasets using the following steps:

\noindent \textbf{user rate filtering}. We only include items that users have rated 5.0 out of 5.0 to capture user purchase performance more accurately.

\noindent \textbf{active user/item filtering}. We select active users and items from all data by including only users who have made more than 20 purchases and items bought by more than 10 users. The statistics of the filtered datasets are shown in Table \ref{table:raw_stats}.

\noindent \textbf{common user filtering}. When we perform CDR tasks, we ensure that the users in the target domain and source domain are the same. This means that we only include the common users and their corresponding purchases.

\noindent \textbf{history sequence length filtering}. We maintain users with source domain history interactions beyond the set threshold. In our experiment, the history sequence length threshold has been set to 20, 30, and 40.


\subsection{Experiment Setting and Metrics}

\begin{table}[t!]
\caption{Number of satisfying common users in cross-domain dataset filtered with different history lengths.}
\scriptsize
\resizebox{0.8\columnwidth}{!}{
\fontsize{9}{10}\selectfont
\begin{tabular}{cc|ccc}
\toprule
\multirow{2}{*}{Source} & \multirow{2}{*}{Target} & \multicolumn{3}{c}{History Sequence Len.} \\
&  & 20 & 30 & 40 \\ \hline
Movies \& TV & Video Games & 439 & 303 & 221 \\
Video Games & Movies \& TV & 439 & 240 & 148 \\
CD \& Vinyl & Movies \& TV & 2127 & 1407 & 1042 \\
Movies \& TV & CD \& Vinyl & 2127 & 1451 & 1050 \\
Video Games & CD \& Vinyl & 197 & 107 & 74 \\
CD \& Vinyl & Video Games & 197 & 126 & 96 \\
Movies \& TV & Electronics & 1172 & 648 & 422 \\
Electronics & Movies \& TV & 1172 & 642 & 391 \\ \hline
\end{tabular}%
}
\label{table:history_length}
\end{table}

\begin{table*}[t!]
\caption{Recommendation performance gain with different candidate list sizes. Metrics with negative performance gains are marked as italics.}
\centering
\resizebox{0.78\textwidth}{!}{%
\fontsize{9}{10}\selectfont
\begin{tabular}{c|ccccccccc}
\toprule
Item Size & H@1 & H@5 & H@10 & P@1 & P@5 & P@10 & N@1 & N@5 & N@10\\\hline
20 & \textbf{54.55\%} & 5.06\% & 10.18\% & \textbf{72.00\%} & \textbf{49.57\%} & 32.43\% & \textbf{54.54\%} & \textbf{23.82\%} & 7.98\%\\
25 & 40.74\% & 5.81\%  & 8.82\% & 51.85\% & 23.99\% & \textbf{35.78\%} & 40.74\% & 13.52\% & \textbf{25.04\%}\\
30 & \textit{-3.33\%} & \textbf{17.54\%} & \textbf{15.07}\% & \textit{-5.71\%} & \textit{-5.37\%} & \textit{-13.20\% }& \textit{-3.33\%} & 1.48\% & \textit{-0.278\%}\\
\bottomrule
\end{tabular}%
}
\label{table:exp_candidates}
\end{table*}

To address research questions RQ1, RQ2, RQ3, and RQ4, we conducted our experiment using OpenAI's GPT-3.5. The results are based on an average of three repeated runs, with the standard deviation reported. Due to budget constraints, if the number of satisfied users exceeds 100 in each experiment, we will randomly select 100 users for evaluation. If the number of satisfied users is below 100, we will evaluate all users. If not specified, the default history sequence length is 30 items, the candidate list size is 20 items, and the negative item sampling strategy is random sampling.

We compared the performance of LLM4CDR using three metrics: HIT@K, MAP@K, and NDCG@K, as summarized in \cite{fayyaz2020recommendation}. In this paper, we use H@K, P@K, and N@K as abbreviations for the metrics.

\noindent \textbf{HIT@K (H@K)}. This determines whether any of the top-K recommended items were present in the test set for a given user.

\noindent \textbf{NDCG@K (N@K)}. NDCG is a widely used metric in information retrieval. It is used to calculate a cumulative score of an ordered set of items.

\noindent \textbf{MAP@K (P@K)}. MAP@K, or Mean Average Precision @ K, is an advanced version of Precision@K.

\subsection{RQ1: How can we wisely select input data?}

Having more historical purchase records generally has a positive impact on recommendation tasks. However, having more abundant historical data often means sacrificing the number of satisfied users, as not all users may have enough purchase records in the source domain. Additionally, including multiple source domains can help alleviate the data sparsity issue to some extent, but it may also introduce other challenges for LLMs, such as understanding the distributions of historical records in the source domain and the relationships across domains. In this section, we examine the impact of input data on the performance of LLM4CDR by evaluating two aspects: the number of satisfied users and LLM4CDR's performance with different lengths of historical sequences.

\subsubsection{Satisfying user count}

This section compares the number of satisfying users across different combinations of source and target domains with source domain history sequence lengths of 20, 30, and 40 items. The results are presented in Table \ref{table:history_length}.

\noindent \textbf{Findings}. The results align with the notion that requiring richer historical purchase data leads to a decrease in the number of satisfying users. This suggests that to reach a broader user base, the threshold for required historical purchase records should be set relatively low. During the prompting stage, more historical data can be included for users with richer purchase records and less data for those with fewer historical interactions.

\subsubsection{Performance with different history sequence lengths}

\begin{figure}[!ht]
    \centering
    \includegraphics[width = 0.48\textwidth]{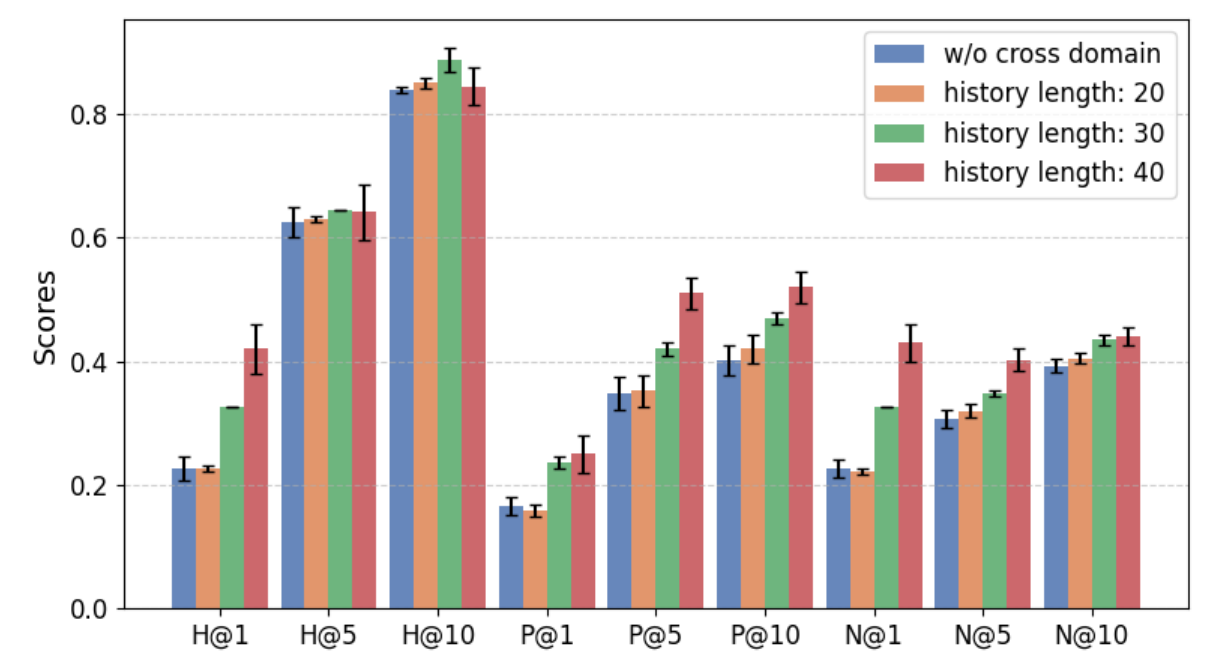}
    \caption{Recommendation performance of LLM4CDR with different source domain history length.}
    \label{fig:exp-history}
\end{figure}

In this section, we compare the performance of LLM4CDR when the history sequence lengths are 20, 30, and 40 items. This comparison is conducted under the condition that the source domain is Video Games, and the target domain is Movies \& TV. The results are shown in Figure \ref{fig:exp-history}.

\noindent\textbf{Findings}. The results show that, in general, LLM4CDR performs better in target domain recommendation when given richer cross-domain information in the user's history. However, when the history sequence is too short (20 items in our case), some metrics such as H@1, P@1, and N@1 perform slightly worse compared to prompting LLMs to perform recommendation tasks without cross-domain information. This could be because insufficient cross-domain history data may provide some transferable knowledge for LLM4CDR to make recommendations, but the predictions may not be confident enough for LLM4CDR to accurately hit the target in the first place. In the following experiments, we set the history sequence length to be 30 items.

\subsection{RQ2: In what way will the candidate list influence LLM4CDR?}

\begin{table}[t!]
\caption{Filtered cross-domain dataset statistics for domain gap influence analysis.}
\centering
\scriptsize
\resizebox{0.98\columnwidth}{!}{
\fontsize{9}{10}\selectfont
\begin{tabular}{c|cccc}
\toprule
CDR Type & Source & Target & \#User & Avg. Len. \\\hline
\multirow{6}{*}{\parbox{1.48cm}{Same \\ Sub-group}}  & Movies \& TV & Video Games & 478 & 59.2 \\
& Video Games & Movies \& TV & 478 & 50.56 \\
 & Movies \& TV & CD \& Vinyl & 743 & 59.2 \\
& CD \& Vinyl & Movies \& TV & 743 & 67.49 \\
 & Video Games & CD \& Vinyl & 157 & 50.56 \\
 & CD \& Vinyl & Video Games & 157 & 67.49 \\\hline
\multirow{2}{*}{\parbox{1.48cm}{Different \\ Sub-groups}}   & Movies and TV & Electronics & 264 & 59.2 \\
 & Electronics & Movies \& TV & 264 & 44.6 \\
\bottomrule
\end{tabular}}%
\label{table:domain-gap}
\end{table}

\begin{table*}[t!]
\caption{Performance comparison when including or excluding source domain history sequence with different domain gap between source domain and target domain.}
\centering
\scriptsize
\resizebox{2\columnwidth}{!}{
\fontsize{9}{10}\selectfont
\begin{tabular}{c|cc|cccccccccc}
\toprule
CDR Type & Source & Target & w/wo Info & H@1 & H@5 &H@10 &P@1 & P@5 & P@10 & N@1 & N@5 & N@10 \\\hline

\multirow{12}{*}{\parbox{1.48cm}{Same \\ Sub-group}}  & \multirow{2}{*}{Movies \& TV}  & \multirow{2}{*}{Video  Games} & wo Info &\textbf{0.2548} &0.6214 &0.8190 &\textbf{0.1885}&\textbf{0.3666}&\textbf{0.4317}&\textbf{0.2548}&0.3094&0.4089\\
&&& w Info & 0.2344&\textbf{0.6675}&\textbf{0.8469}&0.1846&0.3529&0.4167&0.2344&\textbf{0.3158}&\textbf{0.4118}\\\cline{2-13}

& \multirow{2}{*}{Video Games}& \multirow{2}{*}{Movies \& TV} & wo Info&0.2269&0.6250&0.8380&0.1659&0.3485&0.4013&0.2269&0.3075&0.3931\\
&&& w Info & \textbf{0.3272}&\textbf{0.6449}&\textbf{0.8878}&\textbf{0.2361}&\textbf{0.4200}&\textbf{0.4703}&\textbf{0.3272}&\textbf{0.3489}&\textbf{0.4348}\\\cline{2-13}

& \multirow{2}{*}{Movies \& TV}& \multirow{2}{*}{CD \& Vinyl} & wo Info&0.1806&0.5175&0.8302&0.1280&0.2507&0.3129&0.1806&0.2344&0.3434\\
&&& w Info & \textbf{0.2419}&\textbf{0.5968}&\textbf{0.8548}&\textbf{0.1783}&\textbf{0.3262}&\textbf{0.3961}&\textbf{0.2419}&\textbf{0.2870}&\textbf{0.3986}\\\cline{2-13}

& \multirow{2}{*}{CD \& Vinyl}& \multirow{2}{*}{Movies \& TV} & wo Info&0.2460&0.7023&0.8997&0.1996&0.3841&0.4439&0.2460&0.3302&0.4271\\
&&& w Info & \textbf{0.3408}&\textbf{0.7662}&\textbf{0.9058}&\textbf{0.3279}&\textbf{0.5555}&\textbf{0.6109}&\textbf{0.3408}&\textbf{0.4090}&\textbf{0.4955}\\\cline{2-13}

& \multirow{2}{*}{Video Games}& \multirow{2}{*}{CD \& Vinyl} & wo Info&0.1258&0.5346&0.8050&0.1090&\textbf{0.2460}&\textbf{0.2956}&0.1258&0.2288&0.3213\\
&&& w Info & \textbf{0.1509}&\textbf{0.6226}&\textbf{0.8491}&\textbf{0.1373}&0.2370&0.2943&\textbf{0.1509}&\textbf{0.2374}&\textbf{0.3382}\\\cline{2-13}

& \multirow{2}{*}{CD \& Vinyl}& \multirow{2}{*}{Video Games} & wo Info&0.1596&0.5305&0.8498&0.0876&0.2460&0.3233&0.1596&0.2469&\textbf{0.3740}\\
&&& w Info & \textbf{0.1737}&\textbf{0.6009}&\textbf{0.8826}&\textbf{0.1526}&\textbf{0.2896}&\textbf{0.3645}&\textbf{0.1737}&\textbf{0.2516}&0.3709\\\hline

\multirow{4}{*}{\parbox{1.48cm}{Different \\ Sub-groups}}  & \multirow{2}{*}{Movies \& TV} & \multirow{2}{*}{Electronics} & wo Info &\textbf{0.1659}&\textbf{0.5428}&\textbf{0.8241}&\textbf{0.1261}&\textbf{0.2259}&\textbf{0.2867}&\textbf{0.1659}&\textbf{0.2204}&\textbf{0.3318}\\
&&& w Info & 0.1058&0.4333&0.6700&0.0752&0.1698&0.2192&0.1058&0.1753&0.2662\\\cline{2-13}

& \multirow{2}{*}{Electronics}& \multirow{2}{*}{Movies \& TV} & wo Info&\textbf{0.2853}&0.6554&\textbf{0.9040}&\textbf{0.1949}&\textbf{0.3667}&\textbf{0.4410}&\textbf{0.2853}&\textbf{0.3255}&\textbf{0.4429}\\
&&& w Info & 0.2528&\textbf{0.6903}&0.8977&0.1822&0.3552&0.4147&0.2528&0.3112&0.4103\\
\bottomrule

\end{tabular}}%
\label{table:domain-gap-result}
\end{table*}

\begin{table*}[t!]
\caption{Performance comparison of LLM4CDR with different LLMs.}
\centering
\scriptsize
\resizebox{0.98\textwidth}{!}{
\fontsize{9}{10}\selectfont
\begin{tabular}{cc|ccccccccccc}
\toprule

Source & Target & Model & H@1 & H@5 &H@10 &P@1 & P@5 & P@10 & N@1 & N@5 & N@10 & \%imp\\\hline

\multirow{4}{*}{Movies \& TV}  & \multirow{4}{*}{Video Games} & wo Info &\textbf{0.2548} &0.6214 &0.8190 & \textbf{0.1885} &0.3666&0.4317& \textbf{0.2548} &0.3094&0.4089 & \\

&& Ollama &0.2167 &0.5976 &0.7976 &0.1444&0.2902&0.3393&0.2167&0.2880&0.3691 & 0\% \\

&& GPT-3.5 & 0.2344&0.6675&0.8469&0.1846&0.3529&0.4167&0.2344&0.3158&0.4118 & 44.44\% \\

&& GPT-4 & 0.2429&\textbf{0.7167}&\textbf{0.8929}&0.1603& \textbf{0.3790}& \textbf{0.4628}&0.2429& \textbf{0.3445}& \textbf{0.4619} & 66.67\%\\\hline

\multirow{4}{*}{Video Games}  & \multirow{4}{*}{Movies \& TV} &  wo Info&0.2269&0.6250&0.8380&0.1659&0.3485&0.4013&0.2269&0.3075&0.3931 & \\

&& Ollama & 0.1745 & \textbf{0.6897} & 0.8451 &0.1405 &0.3123 &0.3631& 0.1745& 0.3466& 0.4328 & 44.44\%\\

&& GPT-3.5 & \textbf{0.3272}&0.6449&0.8878&\textbf{0.2361}&\textbf{0.4200}&\textbf{0.4703}&\textbf{0.3272}&\textbf{0.3489}&\textbf{0.4348} & 100\%\\

&& GPT-4 & 0.2100& 0.6484 & \textbf{0.8950} & 0.1187 & 0.3256 & 0.3880 & 0.2100 &	0.3170	& 0.4153 & 44.44\%\\\hline

\multirow{4}{*}{Movies \& TV}  & \multirow{4}{*}{CD \& Vinyl} &  wo Info&0.1806&0.5175&0.8302&0.1280&0.2507&0.3129&0.1806&0.2344&0.3434 & \\

&& Ollama & 0.2252	& 0.6783 & 0.8954	& 0.2167 & 0.3423	& 0.4008	& 0.2252	& 0.3085	& 0.4099 & 100\%\\

&& GPT-3.5 & 0.2419& 0.5968& 0.8548& 0.1783 & 0.3262 & 0.3961 & 0.2419 & 0.2870 & 0.3986 & 100\%\\

&& GPT-4 & \textbf{0.3562} & \textbf{0.7941} & \textbf{0.9150} & \textbf{0.3175} & \textbf{0.5176} & \textbf{0.5741} & \textbf{0.3562}	& \textbf{0.3940}	& \textbf{0.4797} & 100\%\\\hline

\multirow{4}{*}{CD \& Vinyl}  & \multirow{4}{*}{Movies \& TV} &  wo Info&0.2460&0.7023&0.8997&0.1996&0.3841&0.4439&0.2460&0.3302&0.4271 & \\

&& Ollama & 0.2205 &	0.7075 &	0.8718	& 0.2013	& 0.3388 & 0.3966 &	0.2205	& 0.3299	& 0.4273 & 33.33\%\\

&& GPT-3.5 & \textbf{0.3408}&\textbf{0.7662}&0.9058&\textbf{0.3279}&\textbf{0.5555}&\textbf{0.6109}&\textbf{0.3408}&\textbf{0.4090}&\textbf{0.4955} & 100\% \\

&& GPT-4 & 0.2816	& 0.7735	& \textbf{0.9417}	& 0.2470	& 0.4756 &	0.5388	& 0.2816	& 0.3866	& 0.4864 & 100\% \\\hline

\multirow{4}{*}{Video Games} & \multirow{4}{*}{CD \& Vinyl} &  wo Info&0.1258&0.5346&0.8050&0.1090& 0.2460 &0.2956&0.1258&0.2288&0.3213 & \\

&& Ollama & 0.1929 &	0.5926	& 0.7929	& 0.1274	& 0.2504	& 0.2984	& 0.1929	& \textbf{0.2555}	& \textbf{0.3464} & 88.89\%\\

&& GPT-3.5 & 0.1509&\textbf{0.6226}&\textbf{0.8491}&\textbf{0.1373}&0.2370&0.2943& 0.1509& 0.2374& 0.3382 & 77.78\%\\

&& GPT-4 & \textbf{0.1950}	& 0.5220	& 0.7862	& 0.1059	& \textbf{0.2594} &	\textbf{0.3142}	& \textbf{0.1950}	& 0.2408	& 0.3419 & 66.67\%\\\hline

\multirow{4}{*}{CD \& Vinyl}  & \multirow{4}{*}{Video Games} &  wo Info&0.1596&0.5305&0.8498&0.0876&0.2460&0.3233&0.1596&0.2469& 0.3740 & \\

&& Ollama & 0.1502	& 0.5728	& 0.8028	& 0.1213	& 0.2644	& 0.3122	& 0.1502	& 0.2696	& 0.3596 & 44.44\% \\

&& GPT-3.5 & 0.1737 & 0.6009 & 0.8826 & 0.1526 & 0.2896 & 0.3645 & 0.1737 & 0.2516 &0.3709 &100\% \\

&& GPT-4 & \textbf{0.2676}	& \textbf{0.6432} &	\textbf{0.8873} & \textbf{0.1753} &	\textbf{0.3571}	& \textbf{0.4336}	& \textbf{0.2676}	& \textbf{0.3163}	& \textbf{0.4326} &100\%\\

\bottomrule

\end{tabular}}%
\label{table:llm_compare}
\end{table*}

\begin{figure*}[h]
    \centering
    \begin{subfigure}{0.3\textwidth}
        \centering
        \includegraphics[width=\textwidth]{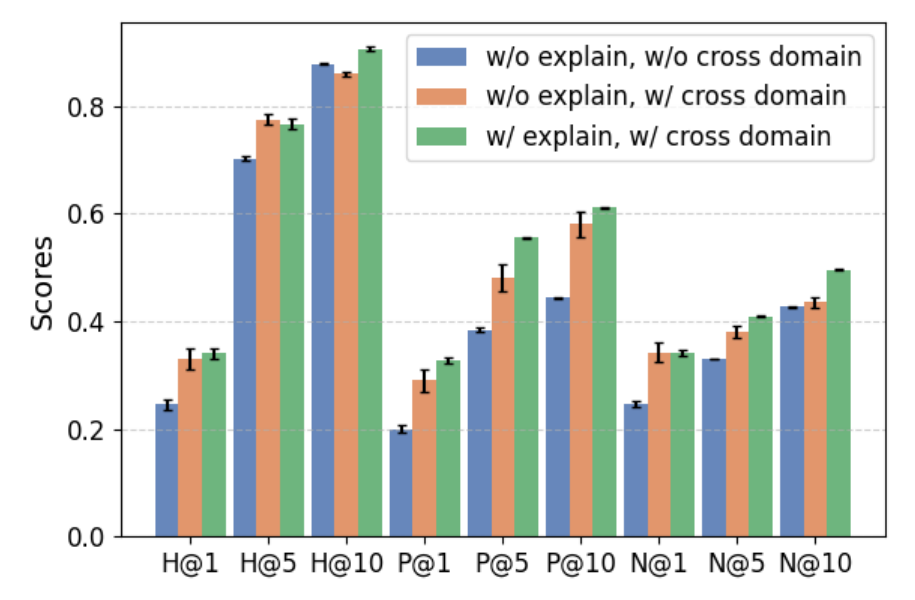}
        \caption{Source domain: CD and Vinyl, Target Domain: Movie and TV}
    \label{fig:random}
    \end{subfigure}
    \hfill
    \begin{subfigure}{0.3\textwidth}
        \centering
        \includegraphics[width=\textwidth]{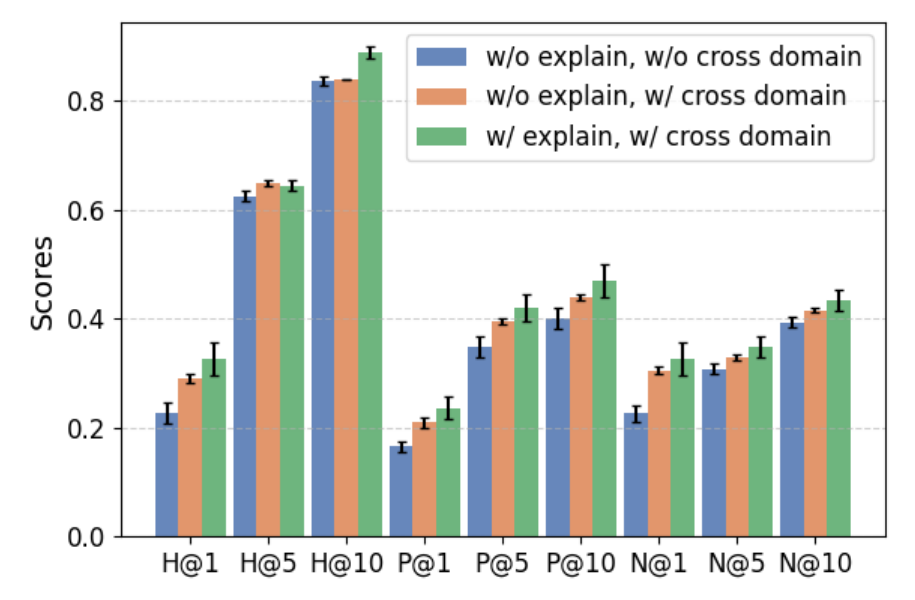}
        \caption{Source domain: Movie and TV, Target Domain: CD and Vinyl}
  \label{fig:similarity}
    \end{subfigure}
    \hfill
    \begin{subfigure}{0.3\textwidth}
        \centering
        \includegraphics[width=\textwidth]{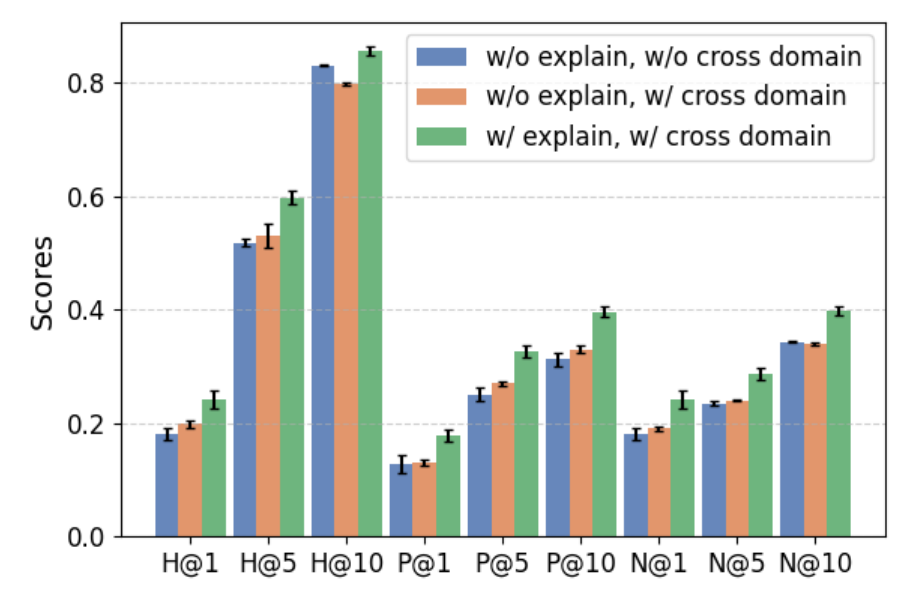}
        \caption{Source domain: Movie and TV, Target Domain: CD and Vinyl}
        \label{fig:popularity}
    \end{subfigure}
    \caption{Performance comparison with/without recommendation guidance.}
    \label{fig:explain-result}
\end{figure*}



We need to investigate how the candidate item set in the target domain affects the overall performance of LLM4CDR since we haven't tested it over the entire candidate item set. We conducted experiments to compare the impact of different candidate list sizes on the CDR performance. We conducted experiments with candidate list sizes of 20, 25, and 30 items. We compared the improvement in metrics when including and excluding the cross-domain history sequence. For our experiments, we designated movie and TV as the source domain and video games as the target domain. The results are presented in Table \ref{table:exp_candidates}.

\noindent \textbf{Findings}. Based on the experiment results, we observed that when the candidate list size is 20 items, most metrics showed the greatest improvement when including the cross-domain history compared to the other candidate item sizes. This suggests that when the candidate list size is relatively small (as in our case, with 20 items), LLM4CDR can effectively leverage cross-domain knowledge to enhance CDR tasks.

\subsection{RQ3: What domain gap can LLM4CDR comprehend?}

In this section, we aim to investigate the conditions under which LLM4CDR can leverage cross-domain information for recommendation tasks. We conduct experiments on cross-domains within the same subgroup or across different subgroups, including subsets of Movies \& TV, Video Games, CD \& Vinyl, and Electronics. The data statistics for this experiment are presented in Table \ref{table:domain-gap}. The performance comparison, with or without the inclusion of the source domain's history sequence, is shown in Table \ref{table:domain-gap-result}.

\noindent \textbf{Findings}. Based on the results, it is evident that when the source domain and target domain belong to the same subgroup, LLM4CDR can effectively utilize cross-domain historical purchase data for recommendation tasks in the target domain. However, in cases of large domain gaps, where the source and target domains are from different subgroups in our experiment, introducing the source domain's historical sequence has a detrimental effect on the overall recommendation performance. One potential reason for this is that when the source and target domains are relatively unrelated, including the source domain's historical sequence introduces irrelevant knowledge to LLM4CDR, which in turn negatively impacts the overall performance.

\subsection{RQ4: Is domain-specific CDR guidance beneficial?}

In this section, we aim to assess whether domain-specific recommendation guidance can genuinely enhance recommendation performance. We compared LLM4CDR's recommendation results with and without recommendation guidance across four cross-domain scenarios. The results are depicted in Figure \ref{fig:explain-result}.

\noindent \textbf{Findings}. The experiment revealed that even without recommendation guidance, the source domain history sequence can still offer valuable cross-domain knowledge for LLM4CDR to make recommendations. However, the improvement in metrics is smaller compared to when using recommendation guidance. This indicates that well-designed recommendation guidance can effectively steer LLM4CDR and enhance the quality of CDR. Therefore, it holds promise to explore effective methods for guiding LLMs in utilizing cross-domain knowledge for recommendation tasks.

\subsection{RQ5: Which LLMs are good at CDR tasks?}

In this section, we are comparing the CDR performance on LLMs with different parameter sizes. We are testing the performance of LLM4CDR with Ollama3, GPT3.5, and GPT-4. After that, we will compare the CDR task performance on different models. In the experiment, the source domain and target domain are from the same subgroup. The history sequence length threshold is 30 items. The results are presented in Table \ref{table:llm_compare}, where $\%imp$ represents the ratio of improved metrics among all metrics compared with the baseline (recommendation performance when no cross-domain information is included).

\noindent \textbf{Findings}. The findings show a positive relationship between the parameter size of LLMs and the performance of CDR. With more parameters, the overall quality of CDR improves. Additionally, it was observed that Ollama3 does not consistently improve performance. This could be due to the complex reasoning process required for CDR tasks, and it is only when the parameter size of LLM exceeds a certain threshold that it can effectively perform CDR tasks.

\section{Conclusion}
\label{conclusion}

We have developed an LLM4CDR pipeline to explore the potential of LLMs in addressing CDR (cross-domain recommendation) problems. In order to thoroughly test the capabilities and limitations of LLM4CDR, we focused on identifying cross-domain scenarios where LLMs excel in transferring knowledge. We also examined the effectiveness of tailoring specific prompts to maximize LLMs' cross-domain recommendation ability. Key findings include: (1) LLM4CDR performs well in cross-domain recommendations when the domain gap is minimal. (2) Providing LLM4CDR with a high-quality recommendation guide, such as specifying common features to consider between the source and target domains, improves its ability to transfer information. (3) LLM4CDR is most effective when the parameter size of LLM is sufficiently large. These findings pave the way for future research endeavors, such as exploring methods to provide LLMs with high-quality CDR guides and efficiently preprocessing known data to better leverage LLMs' cross-domain knowledge transfer capabilities.

\bibliographystyle{ACM-Reference-Format}
\balance
\bibliography{main}

\end{document}